\documentclass[sigconf]{acmart}

\usepackage[utf8]{inputenc}
\usepackage{array}
\usepackage{booktabs}   
\usepackage{caption,subcaption} 
\usepackage{comment}
\usepackage{tikz}
\usepackage{amsmath, array}
\usepackage{diagbox}
\usepackage{graphicx}
\usepackage{tabularx}
\usepackage{parcolumns}
\usepackage{mathtools}
\usepackage{multirow}
\usepackage{enumitem}
\usepackage{centernot}
\usepackage[linesnumbered,ruled,vlined]{algorithm2e}

\usepackage[enable]{easy-todo}
\usepackage{listings}
\usepackage{xspace}
\usepackage{stfloats}
\usepackage[frozencache=true,cachedir=mint,outputdir=.]{minted}
\usepackage{listings, xcolor} 

\definecolor{dkgreen}{rgb}{0,.6,0}
\definecolor{dkblue}{rgb}{0,0,.6}
\definecolor{dkyellow}{cmyk}{0,0,.8,.3}

\definecolor{LightGray}{gray}{0.9}

\usepackage{enumitem}

\renewcommand\footnotetextcopyrightpermission[1]{}
\setlist[description]{style=nextline}

\acmConference[]{}{}

\title{Counterfactual Explanations for Models of Code}
\date{October 2021}

\author{Jürgen Cito}
\affiliation{%
  \institution{TU Wien and Facebook}
  \country{Austria}  
}

\author{Isil Dillig}
\affiliation{%
  \institution{UT Austin$\dagger$\thanks{$\dagger$ Work done at Facebook as visiting scientist}}
  \country{U.S.A.}
}

\author{Vijayaraghavan Murali}
\affiliation{%
  \institution{Facebook}
  \country{U.S.A.}  
}

\author{Satish Chandra}
\affiliation{%
  \institution{Facebook}
  \country{U.S.A.}  
}

\begin{abstract}
Machine learning (ML) models  play an increasingly prevalent role in many software engineering tasks. However, because most models are now powered by opaque deep neural networks, it can be difficult for developers to understand \emph{why} the model came to a certain conclusion and how to \emph{act upon} the model's prediction.
Motivated by this problem, this paper explores counterfactual explanations for models of source code.  Such counterfactual explanations constitute minimal changes to the source code under which the model ``changes its mind". 
We integrate counterfactual explanation generation to models of source code in a real-world setting.
We describe considerations that impact both the ability to find \emph{realistic} and \emph{plausible} counterfactual explanations, as well as the \emph{usefulness} of such explanation to the user of the model.
In a series of experiments we investigate the efficacy of our approach  on three different models, each based on a BERT-like architecture operating over source code.

\end{abstract}

\begin{document}

\maketitle
\newcommand{\model}{\mathcal{M}}
\newcommand{\prog}{P}
\newcommand{\perturb}{\pi}
\section{Introduction}


With the rapid advances in deep learning over the last decade, complex machine learning models have increasingly made their way into software engineering. In particular, models based on deep neural networks are now routinely used in code-related tasks such as bug detection~\cite{deepbugs},  auto-completion~\cite{autocomplete}, type inference~\cite{typewriter}, code summarization~\cite{codesummary1, codesummary2}, and more.


While  deep learning models are remarkably accurate and generally quite effective, one growing concern about their adoption is \emph{interpretability}.  In particular, because deep neural networks are extremely complex, it is often very difficult to understand why they make the predictions  that they do. This issue of interpretability is a particularly relevant concern in software engineering since the outputs of machine learning models are routinely used to make predictions about code quality and non-functional properties of code.  Even worse, because the predictions of these models are often consumed by software developers to change or improve an existing code base, it is particularly important that developers are able to understand \emph{why} machine learning models make certain predictions.

To provide a more concrete illustration, consider a machine learning model that has been trained to detect whether some piece of code contains a security vulnerability~\cite{securityvulnerability}. If the developer does not understand \emph{why} the model thinks a piece of code is vulnerable, they may fail to take a true vulnerability seriously. Furthermore, even when they do take the model prediction seriously, they might not be able to take remedial action if they cannot understand why something is considered to be a vulnerability.

This paper takes a step towards improving the interpretability of  machine learning models used in code for \emph{practical downstream applications.} Motivated by the shortcomings of existing methods like LIME~\cite{lime}, SHAP~\cite{shap}, and attention-based methods~\cite{vaswani2017attention} in the context of source code, we develop a new technique for generating \emph{counterfactual explanations} for models of code.
Such explanations demonstrate how the model's prediction \emph{would} have changed had the program been modified---or, \emph{perturbed}---in a certain way. We believe that counterfactual explanations are particularly useful in the context of software engineering, as they reveal \emph{not only} \emph{which} parts of the code the model is paying attention to, but they also how to \emph{change} the code so that the model makes a different prediction.
This mode of so-called contrastive reasoning through counterfactuals is also aligned with how humans explain complex decisions~\cite{miller:19}.




In order to generate useful counterfactual explanations for code, we have conducted a formative study involving software engineers using different machine learning models at Facebook. Informed by the findings from this formative study, we then formalize what a counterfactual explanation entails in the context of code and present an algorithm for generating such explanations. At the heart of our technique lies a mechanism for modifying the program such that the resulting code is ``natural". This ability is extremely important because perturbations that result in ``unnatural" code can confuse the model by generating programs that come from a different distribution that the model has \emph{not} been trained on. Thus, such modifications often reveal robustness problems in the model as opposed to yielding useful counterfactual explanations. Furthermore, counterfactual explanations that result in ``unnatural code" are also not useful to developers because they do not provide meaningful insights on what a proper contrastive example looks like that flips the model prediction.  We show an example of this in Section~\ref{sec:context}.





In this paper, we propose using \emph{masked language models} (MLMs) as a way of generating  meaningful counterfactual explanations. Our proposed technique ``masks" a small set $S$ of tokens in the original program and uses an MLM to predict a new set of tokens $S'$ that can be used to replace $S$. This results in a ``natural''-looking perturbation of the original program for which the model can make meaningful predictions, as the perturbed program comes from the same distribution that the model has been trained on. Our method then systematically builds a search space of perturbations and tests whether they result in a change of prediction for the relevant downstream task, and if so, returns it as a counterfactual explanation.

We conduct a series of experiments in the context of three machine learning models at Facebook that are applied to source code diffs: performance regression prediction, testplan screenshot prediction, and taint propagation detection. 
We conduct a user study with 3 software engineers and research scientists at Facebook to see whether counterfactual explanations help them discern true positive from false positive predictions (they can in 86\% of cases), and how useful they think these explanations are (83\% of cases).
We also assess how our explanations aligns with human rationales for prediction provided by domain experts. For our study sample, our generated counterfactual explanations correspond to human rationale in over 90\%.

\paragraph{\textbf{Contributions}}

This paper makes the following contributions:
\begin{itemize}
    \item We present a method of generating counterfactual explanations for ML models that predict certain properties of code or code changes.  These properties are typically checked by humans during code review, so if ML models were to be used in place of humans, an explanation of the model predictions is necessary. 
    \item We present desiderata for counterfactual explanations in the context of source code that are \emph{useful} to the end user who has to consume the predictions of the ML models, as well as the counterfactual explanation offered along with it.
    \item We have applied counterfactual explanations for predictions generated by models for three distinct real-world code quality machine learning models.  We show empirical results on the usefulness of these explanations to the users of these models.  We also show, quantitatively, that counterfactual explanations have a better correspondence with human-generated explanation, compared to a previously presented perturbation-based explanation technique~\cite{martensprovost}.
\end{itemize}

The paper is organized as follows. Section~\ref{sec:context} gives the context of this work in terms of the model family we use, as well as the tasks for which these models are trained.  We walk the reader through examples of counterfactual explanations.  Section~\ref{sec:formative} describes informally the requirements from useful counterfactual explanations; these are based on interactions with prospective users.  Section~\ref{sec:problem-def} tackles the problem of explanation generation formally, and presents our core algorithm.  Section~\ref{sec:results} describes our experimental setting and the results.

\section{Context of this work}
\label{sec:context}

\paragraph*{\textbf{BERT-based models of code}}
Deep-learning architectures drawn from NLP have become commonplace in software engineering~\cite{code2vec, codesummary1, codesummary2}. Among these, recently large-scale language models such as BERT~\cite{bert} have shown state-of-the-art performance in NLP tasks, and correspondingly, the software engineering community has adopted these architectures for code-related purposes, e.g. in CodeBERT~\cite{codebert}.   Since our models fall in this category, we give a very brief background here.

The idea behind BERT (and CodeBERT) is to \emph{pre}-train a sequence model using self-supervision.  This is done by giving a neural network the training task of predicting one or more tokens in the sequence that are purposely masked out.  The network's training objective is to predict the masked-out tokens correctly. This is known popularly as a \emph{masked language model}, or MLM.
The purpose of the MLM pre-training is to learn a generic embedding of a token sequence; as in LSTMs, the hidden state after the last token in a sequence is taken to be the embedding of the entire sequence.  

While pre-training is generic, a second step, called \emph{fine tuning}, is applied to customize the network to a specific task at hand.  Here one uses a supervised dataset (i.e., a language fragment as well as a label that a human associates with the fragment), to become a classifier for a specific task.
For instance, in a sentiment analysis application, a (binary) label could be whether the token sequence reflects a positive sentiment.

\paragraph*{\textbf{Automated code review}}
At companies small and large (including Facebook), there is considerable interest in more thorough and more automated code review with the hope of giving feedback to developers early.   Whenever a developer submits a code commit (aka a ``diff''), a series of automated analyses are carried out in addition to a human code review.  These automated analyses try to find both stylistic inconsistencies, and to the extent possible, functional and non-functional issues in code.    

Complementing traditional code analyses are ML-based analyses for a variety of purposes, and are increasingly being used in software engineering tasks in industrial  practice~\cite{ai-in-se}.
Here we illustrate three of these analyses (models), using a common small example.  Our models are based on BERT, described above.

\begin{listing}[!t]
\begin{minted}
	[
	frame=lines,
	framesep=2mm,
	baselinestretch=1.2,
	bgcolor=LightGray,
	fontsize=\footnotesize,
	linenos,
	startinline
	]{php}
private async function storeAndDisplayDialog(
SomeContext $vc,
SomeContent $content,
-   ): Awaitable<SomethingStoreHandle> {
+   ): Awaitable<SomeUIElement> {
-    $store_handle = await SomethingStore::genStoreHandle($vc);
+    $store_handle = await SomethingStore::genHandle($vc);
+    ... other code ...
+    $store_success = await $store_handle->store(
+      $store_handle,
+      $content,
+    );
-    return $store_handle;
+    return await $store_success->genUIElementToRender();
}
\end{minted}
		\caption{A code change that we use as running example. The added lines are marked with a "+" and the deleted lines with a "-"}
		\label{lst:example}
\end{listing}

Listing~\ref{lst:example} represents a  code change that could be made in a diff (for simplicity, we omit other features such as the commit message). It implements a function that stores a piece of content in a data store and returns a UI dialog element to signify that the operation has been successful.   By convention, the added lines are shown with a "+" and the deleted lines are shown with a "-".  The unchanged lines are shown for context, which is also important for the model to train well.

We now discuss the three downstream tasks while referring to Listing~\ref{lst:example}.
We illustrate several aspects of counterfactual explanations in the first of these tasks, and keep the other two brief.

\paragraph*{\textbf{Performance Prediction.}} 

In the diff shown above, replacing line 6 (an optimized operation for store handles) with line 7 (an unoptimized version for generic handles) could cause a performance problem. Lines 8-11  are also added, which could independently cause performance problems.  It is not known at code review time whether or not this diff will cause a performance regression for sure, because the code has not been deployed in a real environment yet.

In support of an intelligent, automated code review, a BERT-based predictive model (trained on a curated set of past performance regressions) provides an educated guess at whether this diff is a likely performance regression.  In this example, suppose that the  model's prediction is ``yes".
In the current status quo, the developer has to start a complex investigation process that involves hypothesis building, statically inspecting potential culprits, and expensive testing with delta-debugging that involves benchmarks~\cite{baltes:15}. If it then turns out that the machine learning prediction was a false positive, it could lead to frustration, as is well-known from existing experiences with deploying any kind of uncertain program analysis~\cite{christakis:16, sadowski:18}.

\begin{listing}[!t]
\begin{minted}
[
frame=lines,
framesep=2mm,
baselinestretch=1.2,
bgcolor=LightGray,
fontsize=\footnotesize,
escapeinside=@@,
startinline
]{php}
-    $store_handle = await SomethingStore::genStoreHandle($vc);
+    $store_handle = await @\colorbox{red}{SomethingStore::genHandle(\$vc)}@;
                           @\colorbox{green}{SomethingStore::genSimple(\$vc)}@
+    ... other code ...
\end{minted}
\caption{Counterfactual explanation of the model's prediction that the code change in Listing~\ref{lst:example}
will cause a performance regression. The explanation says that if we replace the red part with green, the model will no longer make that prediction. We refer to such replacements as \emph{perturbations.}}
\label{lst:counterfactual}
\end{listing}

To help the developer gain better insight about the model's prediction, our system automatically generates a  counterfactual explanation, as shown in Listing~\ref{lst:counterfactual}. This counterfactual states that had \texttt{genStoreHandle} been replaced with \texttt{genSimple} instead of \texttt{genHandle}, then the model would \emph{not} have predicted a performance regression.  Such feedback is useful to the developer, as it highlights that \texttt{genHandle}  is the likely culprit and allows them to ignore changes to the other parts of the code.  Thus, the part elided as "... other code ..." can be ignored by the developer when trying to figure out how to address this automated feedback.


\paragraph*{Multiple Blame Points}
In many cases, the model's prediction does not depend on a single line of the diff, of even a consecutive fragment of code. Rather, there may be multiple parts of the diff that cause the model to make a certain prediction. For instance, Listing~\ref{lst:multipoint} shows a scenario where the counterfactual involves multiple parts of the diff. For example, if the procedure \texttt{store} relies on an optimization performed by \texttt{genStoreHandle}, then it would make sense that the model's prediction is informed by \emph{both} of these function calls. Observe that techniques based on delta debugging~\cite{simplification} might not be able to pinpoint blame points in non-contiguous pieces of code, for example, due to the "...other code..." in this example.


\begin{listing}[!t]
\begin{minted}
[
frame=lines,
framesep=2mm,
baselinestretch=1.2,
bgcolor=LightGray,
fontsize=\footnotesize,
escapeinside=@@,
startinline
]{php}
-    $store_handle = await SomethingStore::genStoreHandle($vc);
+    $store_handle = await @\colorbox{red}{SomethingStore::genHandle(\$vc)}@;
                           @\colorbox{green}{SomethingStore::genSimple(\$vc)}@
+    ... other code ...
+    $store_success = await @\colorbox{red}{\$store\_handle->store}@(
                            @\colorbox{green}{\$store\_handle->probe}@
+      $store_handle,
+      $content,
+    );
\end{minted}
\caption{A counterfactual with two changes that must be made together for the model to change its prediction.}
\label{lst:multipoint}
\end{listing}


\paragraph*{Preserving in-distribution} As illustrated by the previous two examples, our method produces counterfactuals that look ``natural" --- that is, the replacements proposed by our technique \emph{could} have plausibly occurred in the diff. This is an important aspect of this work, as candidate explanations that are implausible often cause the model to ``change its mind" due to robustness issues in the model. For example, consider the candidate explanation shown in Listing~\ref{lst:ood}. In this case, replacing the return value by \texttt{await 5} yields a non-sensical code snippet that comes from a different distribution than the one our model has been trained on. Thus, the model is \emph{not} expected to make reliable predictions for such code and can \emph{erroneously} predict that the diff (after perturbation) does not have a performance regression.  In this case, the out-of-distribution nature of the perturbation results in an ``explanation" that does not blame the true culprit. We want to avoid such out-of-distribution perturbations.

\begin{listing}[b]
\begin{minted}
[
frame=lines,
framesep=2mm,
baselinestretch=1.2,
bgcolor=LightGray,
fontsize=\footnotesize,
escapeinside=@@,
startinline
]{php}
+    ... other code ...
-    return $store_handle;
+    return await @\colorbox{red}{\$store\_success->genUIElementToRender();}@;
                           @\colorbox{green}{5}@ 
\end{minted}
\caption{A counterfactual based on an out-of-distribution code change.}
\label{lst:ood}
\end{listing}







\paragraph*{\textbf{Testplan Screenshot Prediction.}} 
So far, we highlighted salient aspects of our approach in the context of a model for performance prediction. However, such counterfactuals are more broadly useful across different downstream tasks. Next, we illustrate how explanations can be useful in the context of \emph{testplan screenshot prediction}.

When submitting a diff for review, the developer also has to provide a testplan that details how exactly that particular diff should be tested. The testplan usually includes instructions on how to run automated tests that exercise the code change as well as the results of those tests.  A testplan that has to do with UI functionality should also include a screenshot; otherwise it is not a very good quality test plan.  Calling out poor quality test plans is another check that code review is expected to do.

At Facebook, we use a predictive model for testplan quality that indicates whether a testplan should contain a screenshot of the introduced functionality. Such a model is
trained over a curated dataset of diffs with manually-specified binary labels. 
For instance, in our running example, the model might predict that the testplan should require a screenshot, possibly because the functionality that is being changed involves displaying and rendering UI elements (line 14 in the original example).  Our proposed method also helps justify such testplan-related predictions by producing the counterfactual explanation shown in Listing~\ref{lst:testplan}.  With such an explanation, the developer gets a clear indication that the prediction is related to the call to \texttt{genUIElementToRender}, absolving her from the burden of manually inspecting irrelevant parts of a (potentially very large) diff.



\begin{listing}[!t]
\begin{minted}
[
frame=lines,
framesep=2mm,
baselinestretch=1.2,
bgcolor=LightGray,
fontsize=\footnotesize,
escapeinside=@@,
startinline
]{php}
+    $store_handle = await SomethingStore::genHandle($vc);
+    ... other code ...
+    $store_success = await $store_handle->store(
+      $store_handle,
+      $content,
+    );
-    return $store_handle;
+    return await $store_success->@\colorbox{red}{genUIElementToRender}@(); 
                                  @\colorbox{green}{getValue}@    
    }
\end{minted}
\caption{A counterfactual for testplan screenshot prediction model}
\label{lst:testplan}
\end{listing}

\begin{listing}[t]
\begin{minted}
[
frame=lines,
framesep=2mm,
baselinestretch=1.2,
bgcolor=LightGray,
fontsize=\footnotesize,
escapeinside=@@,
startinline
]{php}
   private async function storeAndDisplayDialog(
      SomeContext $vc,
      SomeContent @\colorbox{red}{\$content}@,
       @\colorbox{green}{\$count}@
      ...
+    ... other code ...
+    $store_success = await $store_handle->store(
+      $store_handle,
+      @\colorbox{red}{\$content}@,
        @\colorbox{green}{\$count}@
+    );
\end{minted}
\caption{A counterfactual for taint propagation detection model}
\label{lst:taint}
\end{listing}

\paragraph*{\textbf{Taint Propagation Detection.}} Next, we discuss the usefulness of counterfactuals in the context of a model that predicts whether a diff  introduces data flow that may propagate tainted information. Going back to our running example, a model trained for this task might flag the diff  as suspect because the added code stores some content in a data store (line 9-12 in the original code example).  However, given that the diff makes many other changes, it may not be apparent to the developer why the model makes this prediction. Our technique could also be useful in this context by producing the  counterfactual shown in 
Listing~\ref{lst:taint}. This explanation informs the developer that the model based its decision on the use of the variable name \texttt{content}.



\section{Desiderata for Counterfactual Explanations for Code}
\label{sec:formative}

The series of examples presented in Section~\ref{sec:context} hopefully convey the idea that the results of  ML models can be made more useful, persuasive, and actionable if accompanied by a counterfactual explanation. However, when designing explanatory systems, there are several criteria and pitfalls to consider, some of which have been discussed in the literature~\cite{pitfalls}.  Additionally, since our method  targets models of code, we need to take into account specific considerations of  software engineers submitting their changes for code review. 
For this purpose, we conduct a formative study with three software engineers who are domain experts in the downstream tasks that we investigate in this paper. Our goal is to understand the desiderata of explanation systems as part of the software engineering process.

In this study, we present users with the predictions of the model for specific tasks and  ask them to provide  human rationale for each prediction. We also show them different counterfactual explanations and ask them for feedback. 
Overall, we discuss around 10 predictions with each participant.
At this point of our research, we were mostly interested in open-ended, qualitative feedback to understand the frustrations and confusions of software engineers and assess overall room for improvement. 
This study informed our design decisions that in turn influenced our problem formulation, implementation, and experiments.  In what follows, we briefly summarize our findings, many of which are consistent with 
 similar findings in different (non-code related) contexts.

\paragraph{\textbf{Plausibility and actionability.}}

Plausibility and actionability are regularly mentioned terms in the literature on counterfactual explanations~\cite{face, cf_survey1} and algorithmic recourse~\cite{karimi:21, rawal:20}.
Plausibility  means that we should not generate counterfactuals that are neither realistic nor believably part of the original input.
Similarly to  plausibility constraints in other domains, a code counterfactual is plausible if the code retains its naturalness as a result of the modification.  We showed 
Listing~\ref{lst:counterfactual} as a plausible counterfactual, and Listing~\ref{lst:ood} as one that is \emph{im}plausible.


Another concern frequently brought up in our study is that of \emph{actionability}, i.e., does the explanation show actionable recourse options? 
Pragmatically,  we could take into account actionability constraints in our problem formulation and only allow counterfactuals that modify actionable features of the input. For example, in the case of code diffs, we could restrict modifications to be only applied to the added lines in the diff.
However, after more discussions with our formative study participants,  we realized that  deleted lines and context matter just as much to the model --- otherwise we could have trained the model only on added lines in the first place, making it a lot more myopic and consequently imprecise.
Therefore we decided to \emph{not} restrict which parts of the input to perturb, with the rationale that the model designer can always later apply a filter to the generated counterfactuals in a post-processing step. Listing~\ref{lst:taint}  showed a counterfactual in which one of the perturbations is outside the lines changed in the diff.

\paragraph{\textbf{Consistency.}} Another important finding from our formative study is that counterfactuals that modify \emph{different occurrences} of the \emph{same} identifier are neither particularly useful nor plausible. For example,  consider a diff involving a class called ``SomethingStore". If the modification renames this class in the import statement but not elsewhere in the code, the modified code neither makes sense nor leads to useful counterfactuals. Based on this finding, we decided to restrict code modifications we consider when generating counterfactuals to those that are \emph{consistent}. That is, our approach only allows input modifications that change the same identifier consistently throughout the code diff.




\paragraph{\textbf{Proximity.}}
Multiple formative study participants brought up that the distance between multiple perturbations matters to the comprehensibility of explanations.
Perturbations that were part of the explanation of one instance, but that were ``too spread out" were confusing and not really helpful (e.g., an explanation that added perturbations in the beginning of the diff and at the far end of the diff). We use this finding to design a heuristic to select between multiple candidate counterfactuals. In particular, when our method finds multiple explanations, we prefer those where the modifications are not too far apart.





\section{Counterfactual Explanations for Models of Code}

In this section, we formalize the problem of  generating counterfactual explanations for code and describe our algorithm. 



\subsection{Problem Formulation}\label{sec:problem-def}

 Our problem formulation utilizes the concept of \emph{perturbation} which is a transformation $\perturb$ that can be applied to a program. Given a program $\prog$ and perturbation $\perturb$, we write $\perturb(\prog)$ to denote the resulting program obtained by applying $\perturb$ to $\prog$.

Given a machine learning model $\model$ trained on set $S \sim \mathcal{D}$ for some task $t$ and a program $\prog \sim \mathcal{D}$, a \emph{counterfactual explanation}  is a  perturbation $\perturb$ of $\prog$ satisfying the following criteria:
\begin{enumerate}
    \item The model makes different predictions for the original program and the perturbed version, i.e.,  $\model(\prog) \neq \model(\perturb(\prog))$
    \item The ground truth for $\prog$ and $\pi(\prog)$ are different for task $t$, i.e., $\mathcal{G}(P) \neq \mathcal{G}(\pi(P))$
    \item The perturbed program $\pi(\prog)$ is ``natural", meaning that it comes from the same distribution as the original program $\prog$ (i.e., $\perturb(\prog) \sim \mathcal{D})$
\end{enumerate}

Note that our definition of counterfactual explanation is quite similar to \emph{adversarial examples}~\cite{yefet:20, semanticrobustness} in that both are defined as small perturbations $\perturb$ that cause the model to ``change its mind"  (i.e., condition (1) from above). However, there are two key differences between counterfactual explanations and adversarial examples that are outlined in conditions (2) and (3) in the above definition. First, adversarial examples are inputs that are designed to intentionally fool the model -- that is, from a human's perspective, the prediction result for $\prog$ and $\pi(\prog)$ should be exactly the same. In contrast, a perturbation only makes sense as a counterfactual explanation if $\prog$ and $\pi(\prog)$ are semantically different for task $t$ from the user's perspective. That is, the ground truth prediction for $\prog$ and $\pi(\prog)$, denoted by 
$\mathcal{G}(P)$ and $\mathcal{G}(\pi(P))$ resp.,
should be different, as stipulated in condition (2) of our definition.

In addition, a second key difference between counterfactual explanations and adversarial examples is whether they are ``natural". In particular, since adversarial examples are  specifically crafted to fool the model, there is no requirement that they are drawn from the same distribution that the model has been trained on. On the other hand, for a perturbation $\perturb$ to make sense as an explanation of the model's behavior as opposed to a \emph{robustness} issue, the perturbed program needs to be drawn from the same distribution as the training data, as stipulated in condition (3).

\begin{figure}
\includegraphics[width=0.8\columnwidth, clip]{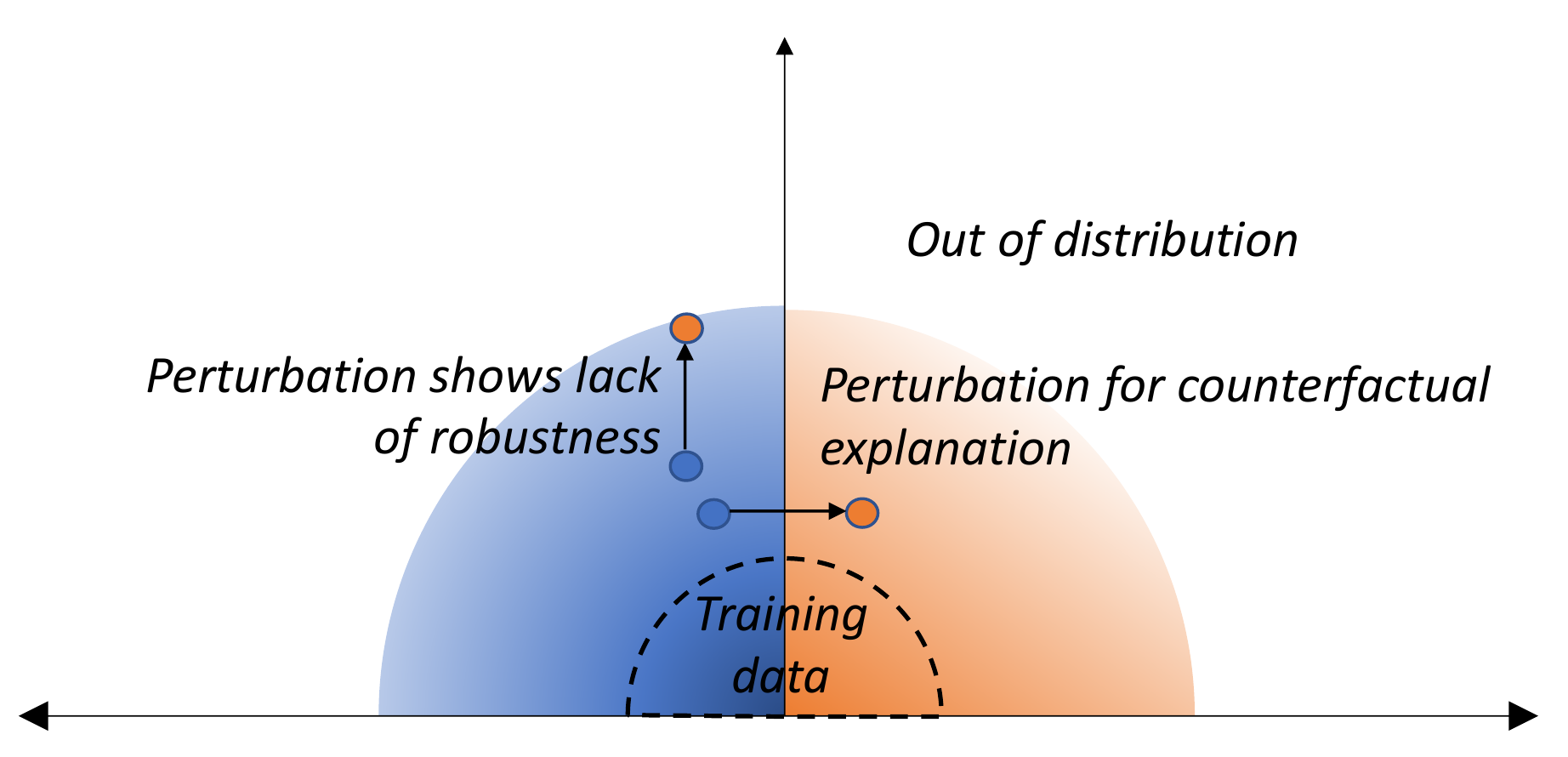}
\caption{The two colors show different classes. As we go out from the center, the data gets more out of distribution. Two perturbations are shown: one that exposes a robustness issue, and the other that is useful for counterfactual explanation. The
colors of the small circles show the class that the model predicts.}
\label{fig:robustvscounterfactual}
\end{figure}

We illustrate the differences between counterfactuals and adversarial examples in Figure~\ref{fig:robustvscounterfactual} in the context of a classification problem with four classes, depicted by different colors.  As we go out from the center, the data gets increasingly out of distribution.  If we want to show an adversarial example, we try a perturbation that pulls the input out of distribution to make the model mispredict, but we want to retain the same ground truth.  If we want to find a counterfactual, we try to perturb such that the ground truth label does change (as well as the model prediction) while staying within distribution.

\paragraph{\textbf{Discussion.}} While our problem formulation stipulates that a counterfactual should flip the \emph{ground truth} prediction, our algorithm for generating counterfactuals can never enforce condition (2), as we do not have access to the ground truth. In fact, an explanation generated by \emph{any} technique (including ours) could violate condition (2). 
However, \emph{assuming} that the model is quite accurate for inputs drawn from the target distribution (i.e., $P \sim\mathcal{D}$), we  have
$\mathcal{M}(P) \approx \mathcal{G}(P)$, so conditions (1) and (2) become interchangeable.  Thus, for models with  high accuracy, an algorithm that ensures conditions (1) and (3) is unlikely to generate counterfactual explanations that violate condition (2). 

Additionally, the contrapositive of the above observation suggests a way of using counterfactuals to diagnose a model's mispredictions. In particular, suppose that an algorithm produces a counterfactual  $\pi$ where the user thinks $\mathcal{G}(P)$ should be the same as  $\mathcal{G}(\pi(P))$. In this case, assuming that the counterfactual generation algorithm enforces conditions (1) and (3), the above assumption that $\mathcal{M}(P) \approx \mathcal{G}(P)$ is violated. Thus, in practice, we have observed a strong correlation between ``non-sensical" counterfactuals and model mispredictions. Said differently, if the algorithm produces a counterfactual that does not make sense to users, this can serve as a strong hint that the model's prediction is wrong.

As an example of this way of diagnosing model mispredictions, revisit Listing~\ref{lst:example}. Suppose, hypothetically, that Listing~\ref{lst:testplan} were offered as the (only) counterfactual for the performance model's prediction of a potential  regression. Suppose further that a human judge immediately sees that the unperturbed and the perturbed code should have the same ground truth, because the UI-related call has nothing to do with performance.  In that case, we can conclude the model's prediction of a performance regression was a misprediction.



\subsection{MLM-Based Perturbations for Code}\label{sec:mlm}

Our problem formulation does not specify the exact nature of perturbations; however, a counterfactual generation algorithm needs to consider a universe of possible modifications to a given code snippet.
To ensure that our counterfactual explanations satisfy the naturalness criterion  (i.e., condition (3)), we propose generating perturbations using \emph{masked language modeling (MLM)}~\cite{bert}. At a high level, the idea is to replace each token with a blank (``mask") and use MLM to come up with a plausible replacement for the original token.  In more detail, our approach  first tokenizes the input program  $\prog$ and produces an indexed sequential representation $\mathcal{T}(\prog) = \{p_i\}_{i=1}^{N}$. Then, given a specific token $p_j$ that we want to perturb, we produce a new sequence  $\mathcal{T}(P') = \langle p_1, \ldots, p_{j-1},$ \texttt[MASK] $, p_{j+1}, \ldots p_N \rangle$ and use MLM to produce a probability distribution for  possible instantiations of \texttt[MASK]. Since MLM leverages all other tokens in the sequence to make predictions,  it can produce more natural perturbations than other (more traditional) language models  that  only leverage preceding tokens in the sequence.

\subsection{Algorithm for Generating Counterfactuals}

We now describe our algorithm for generating counterfactual explanations. Our algorithm takes as input a program $\prog$ and a machine learning model $\mathcal{M}$ and returns a set of counterfactual explanations $E$. Each explanation $e \in E$ is a mapping from a token $t_i \in \prog$ to a new token $t_i'$ such that replacing every $t_i$ with $t_i'$ causes the model $\mathcal{M}$ to change its prediction. 

At a high level, our algorithm explores explanations of increasing size, where the size of an explanation is the number of replaced tokens. Specifically, Algorithm~\ref{alg:generatecounterfactual} maintains a set (called \emph{explore}) of \emph{candidate replacements}, where each candidate is a set of tokens to be replaced. Then, in each iteration, it grows the most promising candidate replacement by one additional token (Line 6) to generate a new set of candidate replacements called \emph{new\_candidates} (Line 7).  Then, for each candidate $c$ in this set,  Algorithm~\ref{alg:exists_cf} invokes a procedure called \textsf{FindCounterfactual} to test whether there exists a counterfactual explanation whose domain is exactly $c$. If so, all explanations returned by \textsf{FindCounterfactual} are added to set \emph{explanations} (line 11). On the other hand, if there is no counterfactual whose domain is $c$, we add $c$ to set \emph{explore} so that the domain of the explanation is grown in subsequent iterations (line 13). Note that, since smaller counterfactual explanations are always preferable, there is no point in adding $c$ to \emph{explore} if  \textsf{FindCounterfactual} returns a non-empty set.

Algorithm~\ref{alg:exists_cf} shows the implementation of the \textsf{FindCounterfactual} procedure. At a high level, this method uses the masked language modeling technique described in Section~\ref{sec:mlm} to find plausible replacements for \emph{all} tokens in $c$ (line 4). That is, given a set of tokens $p_1, \ldots, p_n$, we replace each $p_i$ in this set with a mask and ask the language model to generate a set $S$ of plausible replacements for the tokens $p_1, \ldots, p_n$. Then, for each replacament $\langle p_1', \ldots, p_n'\rangle$ in set $S$, we check whether  substituting each $p_i$ with $p_i'$ \emph{actually} causes the model $\mathcal{M}$ to change its prediction (line 7). If so,  we then add the mapping $[p_1 \mapsto p_1', \ldots, p_k \mapsto p_k']$ to the set of generated counterfactual explanations (line 8).



\begin{algorithm}[t]
  \SetKwInOut{Input}{Input}
  \SetKwInOut{Output}{Output}
\DontPrintSemicolon

  \Input{Input program $P$, Model $\mathcal{M}$: $\mathcal{D} \mapsto \{T, F\} \times \mathcal{R}$ returning a classification and a score, iteration count \texttt{ITER}}
  
  \Output{List of counterfactuals that lead to prediction changes}
  \BlankLine
  
  $explore =\varnothing$\\
  $tokens = \{ t_i~|~t_i \in \mathcal{T}(P) \}$\\
  explanations = $\varnothing$ \\ 
  \For{\_ in \texttt{ITER}+1} {
    $c_{best} = \mathsf{Choose}(explore)$\\
    new\_candidates = $\{ (c_{best}, t_i)~|~t_i \in (tokens \setminus explanations) \}$\\
    \For{$c$ in new\_candidates}{
       
        $E = \mathsf{FindCounterfactual}(P, \mathcal{M}, c)$\\  
        \If{$E \neq \varnothing$} {
            explanations.add($E$)
        }
        \Else {
            explore.add($c$)
        }        
    }
    return explanations
  }

\caption{GenerateCounterfactualExplanations}
\label{alg:generatecounterfactual}
\end{algorithm}

\begin{algorithm}[t]
  \SetKwInOut{Input}{Input}
  \SetKwInOut{Output}{Output}
    \DontPrintSemicolon
  
  \Input{Input program $P$, Model $\mathcal{M}$, Token or token combination to perturb $\langle p_1, \ldots, p_k \rangle$ where $p_i \in \mathcal{T}(P)$\\}
  \Output{Whether counterfactual exists (boolean), Score after perturbation, Counterfactual $P'$ (input after perturbation)}
  \BlankLine
    $E = \varnothing$ \\
    $S= \mathsf{MLM}(P, \langle p_1, \ldots, p_k \rangle)$\\
     \For{$\langle p_1', \ldots, p_k' \rangle \in S$} {
    $P' = P[p_1'/p_1, \ldots, p_k'/p_k]$ \\
    \If{$\mathcal{M}(P) \neq \mathcal{M}(P')$ } {
            $E = E \cup [p_1 \mapsto p_1', \ldots, p_k \mapsto p_k'] $
        }
     }
     return $E$

\caption{FindCounterfactual}
\label{alg:exists_cf}
\end{algorithm}

\section{Experiments}
\label{sec:results}

To explore the effectiveness of our proposed approach, we perform a series of experiments based on models and tasks in the context of  a large deployment scenario at Facebook. In what follows, we provide a brief description of the underlying setting, including  machine learning models and downstream tasks, and then describe the experiment methodology to answer our research questions.

\subsection{Setting}

The models and tasks we employ in our experiments take as input a diff (e.g., akin to pull-requests in open source projects) and assign a task-dependent probability about the code changes (context, additions, deletions) of that diff. Each model has been pre-trained on millions of diffs at Facebook (with no particular downstream task in mind) and then fine-tuned for specific  tasks.









To evaluate our approach, we focus on the downstream tasks described in Section~\ref{sec:context}, namely  performance regression, test plan quality (whether a screenshot is warranted), and detecting whether there is a taint propagation issue.

\subsection{Research Questions and Methodology}

The goal of our experimental evaluation is to address the following research questions.

\paragraph{\textbf{RQ1: How often do users find counterfactual explanations for models of code helpful?}}
We conduct a study with 3 software engineers and research scientists (we will collectively call them ``users" going forward) from different stakeholder teams within Facebook. We randomly sample 30 instances from the validation dataset used during training of these models.
For each instance, we produce counterfactual explanations using our approach.  We then ask the participants whether they found the explanation to be useful to understand the prediction.

\paragraph{\textbf{RQ2: How do users utilize counterfactual explanations to discern between true-positive and false-positive predictions in models of code?}}

We also ask the study participants to assess whether they think prediction is a true-positive or false-positive. They follow a think-aloud protocol in which they are encouraged to verbalize their thought process such that we can capture the rationale of their decision. We qualitatively analyze their responses and report on their experience with utilizing explanations.
    
\paragraph{\textbf{RQ3: Are counterfactual explanations for models of code aligned with human rationales provided by domain experts?}}
    
We perform a case study based on an internal dataset where a team of domain experts for the \emph{taint propagation detection} task had provided human rationales for code changes that we compare to our generated counterfactual explanations.
We randomly sample a set of 30 diffs and filter out instances where we cannot find counterfactual explanations perturbing at most 5 tokens\footnote{Anecdotally, we can say that perturbing any more than 5 tokens will make an explanation practically useless}. We are eventually left with 11 instances for our analysis.

Since our research questions involve human assessment of the counterfactual explanations, we use up to 5 explanations per diff regardless of the number of explanations generated.  The ones shown to users are chosen based on the model's prediction likelihood for the counterfactual. That is, the lower the probability of the positive label from the model, the bigger was the influence of the perturbation, and the better is the explanation.  

\subsection{Results}

\subsubsection*{\textbf{RQ1: Usefulness}}

Our users found the explanations useful or very useful in 83.3\% (25/30) of cases. Very useful explanations made up 30\% (9/30) of cases.
They only found 16.6\% (5/30) of the explanations not useful (or were indifferent about them).
When analyzing these cases together with rationales given by the participants, we found that this mostly had to do with explanations that introduced irrational perturbations.
There are several kinds of perturbations that were considered irrational in the study.
For instance, it was noted that a perturbation did not make sense because a method invocation on a class was perturbed into method that does not exist as part of that class.
A particular explanation was considered not very useful because the counterfactual introduced \emph{too many} perturbations. Even though we aim for sparse explanations, this can happen due to our goal of \emph{consistency} (see Section~\ref{sec:formative}.) If we perturb one particular token (e.g., a type) and it occurs multiple times in the input, we perturb all instances of it. 
Perturbations that occurred in non-actionable parts of the input were also initially dismissed and deemed not useful. However, upon reflection, users noted that it is still useful for models where contextual cues (i.e., where a change occurs) is sometimes more important than the change itself.
This reinforces the perspective on feasibility and actionability we observed in our formative study (Section~\ref{sec:formative}). 

\subsubsection*{\textbf{RQ2: Discerning TP vs FP}}
In addition to eliciting usefulness signals, we wanted to observe how our explanations can help discerning whether a prediction is a true-positive or a false-positive. 
Our users were able to accurately determine the underlying ground truth in 86.6\% (26/30) of cases.
We noticed a distinction in how participants came to the conclusion on how to interpret their explanation.
In true-positive instances, participants noted that the explanation guided them to parts of the code that were aligned with their mental model.
More specifically, the explanation reinforced their hypothesis that had been induced through the prediction.
(One participant did note that while this line of reasoning seems sound, it could be prone to confirmation bias.)
In false-positive instances, participants noted that the strongest signal not to trust the prediction was the level of unreasonableness of the explanation. If the explanation pointed to irrelevant parts of the input or introduced an irrational perturbation, it was a sign that the prediction could probably not be trusted.

Overall, our qualitative analysis showed that the counterfactual explanations provided our study participants with intuition and confidence in understanding exactly where the model is picking up its signal. Whether that signal was reasonable or not helped them decide to trust or discard the prediction.

\begin{table}[h!]
\caption{Experiment overview summarizing the quantitative results of RQ1 and 2}
\label{tab:results}
\begin{tabular}{@{}lll@{}}
\toprule
\textbf{User Study}          & \textbf{TP/FP Guess}   & \textbf{Usefulness}                     \\ \midrule
Overall             & 86.6\% Accuracy & 83.3\% useful / 16.6\% not \\
Performance         & 85\% Accuracy  & 85\% useful / 15\% not \\
Testplan Screenshot & 90\% Accuracy & 80\% useful / 20\% not \\ 
\bottomrule
\end{tabular}
\end{table}



\subsubsection*{\textbf{RQ3: Correspondence with Human Rationale}}

We investigate to what extent generated explanations align with rationale provided by human experts.
To determine this alignment, two of the authors compare the generated explanations to the rationales.
Specifically, we map the natural language description to tokens of the source code.
We consider an explanation to match the rationale if their token attributions overlap at least 50\%.
In our sample, $\sim$90\% (10/11) of the instance explanations (at least one of the top 5 offered) aligned with the human rationale.  Even in the one case we deemed to not align entirely, while the top 5 explanations by our scoring did not align with the human rationale, two among the top 10 did.

In addition to analyzing alignment, we wanted to see how our approach compares to a representative baseline from the literature.
Thus, we also generate explanations that are based on occlusion (i.e., token removal instead of replacement) using the SEDC algorithm~\cite{martensprovost}, a greedy approach that successively removes token combinations from the input until the prediction flips.
Occlusion has been used as a simple, yet effective, baseline in related work in NLP~\cite{hotflip, li_nlp:16, feng:18}.

Table~\ref{tab:quantresults} provides a quantitative overview on the results of our approach (CFEX) with the baseline (SEDC) including: number of explanations generated (\textbf{\#Exp}), size of the diff measured in number of tokens (\textbf{Size}), and summary statistics on the size of explanations in number of tokens (\textbf{Avg, Min, Max}).
We compare both approaches by determining a winner (\textbf{Wins}) based on the following criteria: First, the approach needs to produce an explanation that aligns with the human rationale. If both approaches generate aligning explanations, the shorter explanation wins (less number of attributive tokens).
If the explanation size is the same, it is considered a tie.
In our results, we observe wins or ties in $\sim$90\% (10/11) for CFEX and $\sim$19\% (2/11) for SEDC.

This indicates that CEFX produces explanations that are more aligned with human explanation while also producing shorter explanations than the baseline technique does.
Overall, we were able to see that counterfactual explanations generated through perturbations proposed by MLM are \emph{useful}, can help users discern false- and true-positives, and seem likely to align human rationale.


\begin{table}[h]
\caption{Overview of generated counterfactual explanations (CFEX) and explanations generated by token removal (SEDC)}
\label{tab:quantresults}
\begin{tabular}{ll|llllll}
   &      & \textbf{\# Exp} & \textbf{Size} & \textbf{Avg} & \textbf{Min} & \textbf{Max} & \textbf{Wins} \\ \hline
1  & CFEX & 1               & 234           & 4            & 4            & 4            & \textbf{x}    \\
   & SEDC & -               &               &              &              &              &               \\ \hline
2  & CFEX & 7               & 143           & 5            & 5            & 5            & \textbf{x}    \\
   & SEDC & 2               & 143           & 4.5          & 4            & 5            & \textbf{x}    \\ \hline
3  & CFEX & 23              & 274           & 3.83         & 3            & 4            &               \\
   & SEDC & 13              & 274           & 4            & 4            & 4            & \textbf{x}    \\ \hline
4  & CFEX & 11              & 307           & 2.91         & 2            & 3            & \textbf{x}    \\
   & SEDC & 11              & 307           & 3.64         & 1            & 4            &     \\ \hline
5  & CFEX & 8               & 48            & 2.75         & 2            & 3            & \textbf{x}    \\
   & SEDC & 2               & 48            & 3            & 3            & 3            &     \\ \hline
6  & CFEX & 75              & 315           & 4            & 4            & 4            & \textbf{x}    \\
   & SEDC & -               &               &              &              &              &               \\ \hline
7  & CFEX & 13              & 96            & 3.62         & 3            & 4            & \textbf{x}    \\
   & SEDC & -               &               &              &              &              &               \\ \hline
8  & CFEX & 27              & 219           & 2.96         & 2            & 3            & \textbf{x}    \\
   & SEDC & 22              & 219           & 3.95         & 3            & 4            &               \\ \hline
9  & CFEX & 88              & 292           & 2.77         & 2            & 3            & \textbf{x}    \\
   & SEDC & 30              & 292           & 3            & 3            & 3            &               \\ \hline
10 & CFEX & 124             & 301           & 2.95         & 2            & 3            & \textbf{x}    \\
   & SEDC & -               &               &              &              &              &               \\ \hline
11 & CFEX & 15              & 117           & 2.27         & 1            & 3            & \textbf{x}    \\
   & SEDC & 22              & 117           & 2.23         & 1            & 3            &              
\end{tabular}
\end{table}

\section{Discussion}



\subsubsection*{\textbf{Deployment Considerations}}

While most research on explanatory systems discusses how to provide faithful explanations to predictions, they implicitly require that prediction to be a true-positive. Most research papers discussing explanations do not even consider false-positives in their evaluation scenarios.
However, in the context of deploying explanations as part of large-scale machine learning deployments, considering false-positives is vital to the success of the model. This is especially true in the context of learned models for code quality, where developers are rightly disillusioned with the lack of certainty and soundness of developer tools, leading them to chase ghosts in the light of wrong predictions.
That is why we put an emphasis on evaluating our approach not only on a subjective consideration of ``usefulness", but also whether (and how) an explanation instills trust in both correct and wrong predictions.
Another aspect we want to consider for deployment is enabling interactive use with our generated counterfactuals to allow for exploration of decision boundaries (Figure~\ref{fig:robustvscounterfactual}).





\subsubsection*{\textbf{Interactive Counterfactual Exploration}}

Using generative models to produce counterfactuals incurs non-negligeble runtime costs that ranges in the order of 30 seconds for smaller diffs to up to 10 minutes on the very large side of the diff size spectrum.
In our current setting (Section~\ref{sec:context}), we can allow ourselves to process counterfactual generation offline and attach it as part of the code review.
However, we envision counterfactual explanations to become part of an interactive exploration process.
While part of making such an interactive experience possible is certainly performance engineering, we may have to think of other creative ways to make counterfactual generation with language models more instant.
A possibility we want to explore is leveraging the traces of token replacements produced in the offline search to learn a neural model that mimics the MLM filling with much faster inference times.

\subsubsection*{\textbf{Limitations and Scope}}

We developed and applied the proposed approach in the context of workflows and tools within Facebook. While nuances of the internal workflow may have specific peculiarities, we generally think that the mode of submitting commits for code review is widely established both in wider industry and open-source projects.
Also, while we run our experiments on large transformer, i.e. BERT-like, models, our approach and algorithm are model agnostic and only require repeated access to label and score information that most statistical learning models provide.
Nevertheless, other kinds of models could exhibit other characteristics when it comes to the kinds of counterfactuals that are produced.

\section{Related Work}

There has been significant recent interest in improving the interpretability of machine learning models. Efforts in this space can be broadly classified into two categories, namely \emph{local} and \emph{global} interpretability. Local explanability techniques aim to provide justification for predictions on a specific input~\cite{lime,shap,anchors}. On the other hand, global explanability techniques try to explain the behavior of the \emph{whole} model, for instance by constructing a simpler surrogate model that emulates the original model~\cite{global1,global2}. Since our goal in this work is to provide justifications for individual predictions, this work falls under \emph{local explanation} techniques. In what follows, we give a more detailed overview of relevant work in this space. 



\paragraph{White-box techniques for local explanations} Techniques for generating local explanations can be further classified as being either \emph{white-box} or \emph{black-box}. As their name indicates, white-box techniques are customized to specific ML models and exploit the internals of model when generating explanations. A common approach to white-box interpretability of deep learning is through so-called \emph{attention mechanisms} where the idea is to use weights of attention layers inside the network to determine the importance of each input token
~\cite{clark:19, galassi:20, li:16, vashishth:19}. However, recent work has shown that different weights can lead to the same prediction and has called into question whether attention-based mechanisms are actually meaningful as explanations~\cite{jain:19}. Similarly, other work has  shown that it is possible to systematically manipulate attention while still retaining the same prediction~\cite{pruthi:20}.

Another popular white-box technique for generating local explanations is \emph{integrated gradients}~\cite{sundararajan:17}. The high-level idea behind this method is to create interpolations of the input and evaluate the model on those interpolated inputs. Unfortunately, unlike image data, code does not lend itself easily to such interpolation. For instance, there is no meaningful token that can be obtained by combining the embedding of a pad token and that of a keyword token. In fact, we initially investigated using integrated gradients for generating local explanations for models of code, but we were not successful in generating useful explanations. This is, presumably, because unlike images, there is no natural interpolation between a zero token and the token in the input text.

\paragraph{Perturbation-based explanation mechanisms} Another common approach for generating local explanations is through perturbation-based mechanisms such as LIME~\cite{lime} and SHAP~\cite{shap}. These techniques remove or replace a subset of the features in the input space and track the score differential of the model's prediction. The aggregated score difference over many samples is then attributed to features involved in the perturbation. While these techniques can be applied to multiple types of ML models, they do not generate counterfactual explanations. Instead, they highlight input features that are most important for the model's prediction.

In the realm of understanding for source code models, recent work uses delta-debugging techniques to reduce a program to a set of statements that is minimal and still preserves the initial model prediction~\cite{simplification, suneja:21}. The intuition here is that essentially the remaining statements are the important signal being picked up by the model. Another effort empirically show that attention scores in neural networks are highly correlated with code perturbations (statement removals in the source code input)~\cite{autofocus}. However, these works have not investigated (or considered) the effects of token removal that may lead to out-of-distribution inputs. These in turn can lead to unreliable prediction outcomes~\cite{datasetshift} and misleading attention scores~\cite{jain:19}.

\paragraph{Counterfactual explanations} Another common approach to local interpretability is through the generation of counterfactuals. These techniques are very related to the previously discussed perturbation-based mechanisms, but they come with a stronger guarantee, namely that the perturbations are \emph{guaranteed} to change the model's prediction. The generation of counterfactual explanations has received significant attention in the NLP community~\cite{martensprovost,polyjuice,explain-nlp,generate,generative}. In the simplest case, these counterfactuals can be generated by deleting words from the input text~\cite{martensprovost} or via rewrite-rules such as adding negations  or shuffling words~\cite{polyjuice}. Similar to our goal of generating natural-looking programs, these techniques also aim to generate \emph{fluent} text that is grammatically correct. Among the counterfactual generation techniques, perhaps the most relevant to ours is the \emph{minimal contrastive editing (MiCE)} technique of~\cite{explain-nlp}.  Specifically, they train an \emph{editor} to predict words to edit and use a generative model (called \emph{predictor}) to predict replacements for the chosen words.

\paragraph{Interpretability of SE models.} There has also been recent interest in improving interpretability of models used in software engineering~\cite{simplification,md,autofocus,probing}. Two of these efforts~\cite{simplification,probing} propose to simplify the code while retaining the model prediction. Another effort called AutoFocus~\cite{autofocus} aims to rate and visualize the relative importance of different code elements by using a combination of attention layers in the neural network and deleting statements in the program. Another recent effort~\cite{md} aims for global interpretability and helps model developers by identifying which types of inputs the model performs poorly on. We believe that the counterfactual explanation generation technique proposed in this paper complements all of these efforts on improving SE model interpretability.

\bibliographystyle{ACM-Reference-Format}
\bibliography{references}

\end{document}